\documentclass[aps,pra,reprint,a4paper,amsmath,amsfonts,superscriptaddress,longbibliography]{revtex4-1}

\usepackage[T1]{fontenc}
\usepackage[utf8]{inputenc}
\usepackage{braket}
\usepackage{bm}
\usepackage{graphicx}

\DeclareMathOperator{\dif}{d \!}

\DeclareMathOperator{\SO}{SO}

\DeclareMathOperator{\real}{Re}
\DeclareMathOperator{\imag}{Im}

\newcommand{\od}[3][]{\ensuremath{%
  \frac{\dif{^{#1}}#2}{\dif{#3^{#1}}}}}
\newcommand{\pd}[3][]{\ensuremath{%
  \frac{\partial{^{#1}}#2}{\partial#3^{#1}}}}

\newcommand{\half}{{\ensuremath{\tfrac{1}{2}}}}

\renewcommand{\vec}[1]{\boldsymbol{#1}}

\newcommand{\toyg}{{\mathfrak{g}}}

\long\def\f#1/#2{\frac{#1}{#2}}

\begin{document}

\title{The Goldilocks model of separable, zero-range, few-body interactions in one-dimensional harmonic traps}

\author{Molte Emil Strange Andersen}
\affiliation{Institut for Fysik og Astronomi, Aarhus Universitet, Denmark}

\author{N.L. Harshman}
\affiliation{Institut for Fysik og Astronomi, Aarhus Universitet, Denmark}
\affiliation{Department of Physics, American University, Washington, DC, USA}

\author{Nikolaj Thomas Zinner}
\affiliation{Institut for Fysik og Astronomi, Aarhus Universitet, Denmark}

\date{\today}

\begin{abstract}
This article introduces the ``Goldilocks model'' for a few repulsively interacting particles trapped in a one-dimensional harmonic well and provides exact solutions for the three-particle case. The Goldilocks model shares features with two other well-known systems, the Calogero model and the contact-interaction model, and coincides with them in limiting cases. However, those models have purely two-body interactions whereas this model has intrinsically few-body interactions. Comparing these three models provides clarifying distinctions among the properties of symmetry, separability and integrability. The model's analytic solutions provide a useful basis to improve approximation schemes, especially near the unitary limit of hard-core contact interactions.
\end{abstract}
\maketitle

\section{Introduction}

One challenge of few-body physics is that the degrees of freedom grow more rapidly than the constraints from symmetry. This hard truth impedes many straightforward analytic and numerical approaches to extracting physics even from simple models. The restriction to one-dimensional models generally makes calculations more tractable because the balance of symmetry versus degrees of freedom is more favorable. Sometimes the balance is so favorable that the one-dimensional model is solvable, and it becomes a wellspring for physical and mathematical insight about few-body and many-body dynamical systems. As a result, there is a long and productive history of one-dimensional solvable models in many branches of physics, in particular mathematical and condensed matter physics. One important example is the zero-range, contact-interaction (or delta-interaction) model in one dimension, which includes the Tonks-Girardeau gas \cite{girardeau_relationship_1960}, the Lieb-Liniger bosons model~\cite{lieb_exact_1963}, and its extensions to multicomponent bosons and fermions~\cite{gaudin_bethe_2014}. Another is the Calogero model (also called the Calogero-Moser model)~\cite{calogero1969a,calogero1971} with inverse-square interactions and its numerous generalizations, for example Refs.~\onlinecite{calogero_calogero-moser_2008,olshanetsky_quantum_1983}.

Recently, interest in one-dimensional few-body models has further increased because of ongoing experimental advances with ultracold atoms in effectively one-dimensional optical traps. In these cold atom experiments, the range of interaction is typically much shorter than other length scales, and the system behaves like an effective one-dimensional contact-interaction model whose interaction strength is determined by an interplay between the Feschbach and confinement-induced resonances~\cite{olshanii_atomic_1998}. Additionally, the optical trap is well-modeled as a harmonic potential. Experiments with many-atom cold gases in a trap at near the ``unitary limit'' of hard-core contact-interactions~\cite{kinoshita_observation_2004, paredes_tonksgirardeau_2004, kinoshita_quantum_2006} have demonstrated the importance of integrability for understanding thermalization and non-equilibrium quench dynamics~\cite{rigol_thermalization_2008,PhysRevA.89.033601}. Few-atom experiments with tunable interactions, well shapes and spin mixtures~\cite{serwane_deterministic_2011, wenz_few_2013, murmann_two_2015, PhysRevLett.115.215301} offer exciting possibilities for quantum simulation of condensed matter systems from the ``bottom up''~\cite{murmann_two_2015, Zinner2014, PhysRevA.95.053616}. In the near unitary limit, these systems can be mapped onto one-dimensional spin chains that have coupling constants which depend on the trap shape~\cite{volosniev_strongly_2014, PhysRevA.90.013611, levinsen_strong-coupling_2015, PhysRevA.93.013617,harshman_identical_2017}. The possibility for precision control of these systems has also inspired practical proposals for embodying and processing quantum information in such systems~\cite{PhysRevA.91.023620}. These experiments and potential applications motivate the search for solvable models that allow qualitative analysis and aid quantitative precision of prediction and control.

Towards this end, this article introduces a model for interacting particles in one-dimensional harmonic traps and compares it to the contact-interaction model and the Calogero model.
The Hamiltonian for the model we consider in natural units is
\begin{equation}\label{ham:toy}
H_\toyg= \frac{1}{2} \sum_{i=1}^N \left( - \frac{\partial^2}{\partial x_i^2} +  x_i^2 \right)+ \frac{\sqrt{2}\toyg}{\rho} \sum_{\langle i,j \rangle} \delta(x_i - x_j),
\end{equation}
where  the sum is over all pairs $\langle i , j \rangle$ and $\rho$ is the relative hyperradius defined for $N$ particles as 
\begin{equation}\label{int:Nbody}
\rho = \frac{1}{\sqrt{N}}\sqrt{(N-1) \sum_{i=1}^N x_i^2 -2\sum_{\langle i , j \rangle} x_i x_j}.
\end{equation}
We shall only consider repulsive interactions ($\toyg>0$) to avoid the problem of the wave function `falling to the center' \cite{landau-lifshitz}. For convenience, we will call this the Goldilocks model because of its cozy position between those two other famous models. The Calogero model is too hard (particles cannot transmit past each other) and the contact-interaction is too soft (in the sense that the contact-interaction is too weak at short distances to provide separability and solvability). Also, Goldilocks seems appropriate because the interaction modification is `just right' to make the model analytically solvable with three particles.

For comparison, the Hamiltonian for the contact-interaction model in a harmonic trap is
\begin{equation}\label{ham:con}
H_g= \frac{1}{2} \sum_{i=1}^N \left( - \frac{\partial^2}{\partial x_i^2} +   x_i^2 \right) + g \sum_{\langle i,j \rangle} \delta(x_i - x_j)
\end{equation}
and the Calogero model with a harmonic trap is
\begin{equation}\label{ham:cal}
H_\gamma = \frac{1}{2} \sum_{i=1}^N \left( - \frac{\partial^2}{\partial x_i^2} +  x_i^2\right) + \gamma \sum_{\langle i,j \rangle} \frac{1}{|x_i - x_j|^2}.
\end{equation}
The Calogero model is exactly solvable and integrable for positive $\gamma$ (in fact, it is maximally superintegrable~\cite{PhysRevD.90.101701}). In comparison, for $N>2$ the contact-interaction model with a harmonic trap is only exactly solvable and integrable for no interactions $g=0$ and in the unitary limit $g \to \infty$~\cite{PhysRevLett.99.230402, deuretzbacher_exact_2008, guan_exact_2009,fang_exact_2011, PhysRevA.89.033633, Harshman2016}. When $\toyg \to \infty$, the Goldilocks model Hamiltonian coincides with the contact-interaction Hamiltonian at the unitary limit $g \to \infty$. Further, when $\gamma \to 0$ the Calogero model is also equivalent to the unitary limit of the contact-interaction~\cite{calogero_calogero-moser_2008}.

The Goldilocks model shares different properties with the Calogero and the contact-interaction model.  Like the contact-interaction model (but unlike the Calogero model), particles in the Goldilocks model transmit past each other, except in the unitary limit. On the other hand, like the Calogero model (but unlike the contact-interaction model for $N>2$), the relative hyperradial and hyperangular coordinates separate for this Goldilocks model. As a result, interactions are diffractionless, in the sense of Sutherland~\cite{sutherland_beautiful_2004, lamacraft_diffractive_2013}.  However, unlike both those other models, the interaction in (\ref{int:Nbody}) is an intrinsically $N$-body interaction and might seem peculiar from a physical point of view. For finite interaction strength, the pair-wise interaction is stronger when all $N$ particles are close to each other, and weaker when even just one of the $N$ particles is pulled far away. In order to have hyperradial separability in a harmonic trap, we must have that $V_\text{int} \mapsto V_\text{int} / \alpha^2$ under the transformation $\rho \mapsto \alpha\rho$ for some $\alpha>0$. Only the Calogero model has a potential with this property that is also a sum of Galilean invariant, two-body interactions. The Goldilocks model does possess Galilean invariant $N$-body interactions, and in that way is similar to separable, solvable $N$-body interaction models like the Wolfes model~\cite{doi:10.1063/1.1666826}, the Jain-Khare model~\cite{JAIN199935}, and other truncated Calogero-Sutherland type models~\cite{PhysRevB.95.205135}.

Despite the physical peculiarity of the Goldilocks model, we argue that it is worth attention for several reasons. First, it is difficult to get accurate energies and wave functions for the contact-interaction model at large strengths because the contact-interaction introduces cusps in the few-body wave functions that require high energy scales to accurately capture. As a result, even for three particles methods like exact diagonalization converge slowly unless techniques to address the cusp are used~\cite{lindgren_fermionization_2014}. The three-body solutions for the Goldilocks model could also serve as basis for variational methods or perturbation methods. This would be particularly useful in the near unitary limit, where  renormalization is required~\cite{sen_perturbation_1999, gharashi_one-dimensional_2015}. More generally, comparing these three models shows how symmetry, separability and integrability are distinct but related features. This model shows how the solvability of a model comes down to boundary conditions on surfaces and singular points in configuration space, and this insight can be extended to other models in higher dimensions. Finally (and explicitly hopefully), solvable models end up being useful, often in ways never intended. For example, the Tonks-Girardeau gas was a purely theoretical example for more than fifty years before becoming experimentally realized.

An outline for the rest of the paper follows. In Section II, we give the exact solutions for the three-body version of (\ref{int:Nbody}), which is fully separable in center-of-mass and relative polar coordinates and integrable for any $\toyg > 0$. In section III, we compare these to solutions of the Calogero and contact-interaction three-body models in the non-interacting, weakly interacting, near-unitary, and unitary limits. In Section IV, we consider extensions to more particles and discuss integrability and symmetry for $N \geq 3$. Section V suggests direction for extensions of the model and for future research.

\section{Exact solution for three particles}

First we consider three distinguishable particles and solve the problem through separation of variables. The interaction part of the Goldilocks Hamiltonian (\ref{int:Nbody}) can be rewritten as 
\begin{multline}
  V_\text{int} =\toyg \sqrt{\frac{3}{2}} \bigg( \frac{\delta(x_1-x_2)}{|\half(x_1+x_2)-x_3|} \\
  + \frac{\delta(x_2-x_3)}{|\half(x_2+x_3)-x_1|} + \frac{\delta(x_3-x_1)}{|\half(x_3+x_1)-x_2|} \bigg) \label{eq:interaction-potential}
\end{multline}
to explicitly demonstrate that each pairwise interaction term depends on the distance from the center of mass of the pair to the third ``spectating'' particle.
We convert to relative Jacobi coordinates $\vec{x}'=J\vec{x}$ through the orthogonal matrix
\begin{equation}
  J = \begin{bmatrix}
    \frac{1}{\sqrt{2}} & \frac{-1}{\sqrt{2}} & 0 \\
    \frac{1}{\sqrt{6}} & \frac{1}{\sqrt{6}} & \frac{-2}{\sqrt{6}} \\
    \frac{1}{\sqrt{3}} & \frac{1}{\sqrt{3}} & \frac{1}{\sqrt{3}}
  \end{bmatrix}.
\end{equation}
The motion of the center of mass (with coordinate $x_3'$) separates from the relative motion, and from here on we shall restrict to considering only the relative motion.

We further convert the Jacobi coordinates into hyperspherical coordinates with radius
\begin{equation}
  \begin{aligned}
      \rho
      &=\sqrt{x_1'^2+x_2'^2} \\
      &= \sqrt{\f2/3 \left(x_1^2 + x_2^2 + x_3^2 - x_1x_2 - x_2x_3 - x_3x_1\right)}
  \end{aligned}
\end{equation}
and angle $\tan(\phi)=x_2'/x_1'$.
In these coordinates, the relative Hamiltonian is
\begin{equation}\label{eq:hrel}
  H^\text{rel}_\toyg = \frac{1}{2}\left(-\pd[2]{}{\rho} - \frac{1}{\rho} \pd{}{\rho} + \rho^2 + \frac{\Lambda_\toyg}{\rho^2} \right),
\end{equation}
having defined the operator
\begin{equation}
  \Lambda_\toyg = -\od[2]{}{\phi} + \toyg \sum_{n=1}^6 \delta(\phi-\phi_n). \label{eq:Lambda-defn}
\end{equation}
The delta functions act along the lines $\phi_n = \frac\pi6$, $\frac{3\pi}6$, $\frac{5\pi}6$, $\frac{7\pi}6$, $\frac{9\pi}6$, $\frac{11\pi}6$ for $n=1,\dotsc,6$, respectively. We see that due to the factor of $1/\rho^2$ in the interaction potential, the angular part separates from the radial part of the problem. Note that for finite $g$ the contact-interaction model cannot be put in the form (\ref{eq:hrel}) and does not have hyperradial-hyperangular separability. However, the Calogero model can. The relative Hamiltonian has the same form as (\ref{eq:hrel}), except with $\Lambda_\toyg$ replaced by the angular operator $\Lambda_\gamma$ with the form
\begin{equation}
  \Lambda_\gamma = -\od[2]{}{\phi} + \gamma \frac{9}{2 \cos^2(3\phi)}.
\end{equation}
Note this operator has the same six singular angles $\phi_n = (2n-1)\pi/6$.

Using separation of variables, $\psi(\rho,\phi) = R(\rho)\Phi(\phi)$, the Schr\"odinger equation splits into two eigenvalue problems.
Solving first the angular equation
\begin{equation}
  \Lambda_\toyg \Phi(\phi) = \lambda^2 \Phi(\phi) \label{eq:angular-eigenvalue-problem}
\end{equation}
then allows us to substitute its eigenvalue---the angular quantum number $\lambda$---into the radial equation
\begin{equation}
  E R(\rho) = \frac{1}{2} \left( -\pd[2]{}{\rho} - \frac{1}{\rho} \pd{}{\rho} + \rho^2 + \frac{\lambda^2}{\rho^2} \right) R(\rho).
\end{equation}
The solution to the radial equation is well known:
\begin{equation}
  R_\nu^\lambda(\rho) = \sqrt{\frac{2\nu!}{\Gamma(\nu+\lambda+1)}} \rho^\lambda e^{-\rho^2/2} L_\nu^\lambda(\rho^2),
\end{equation}
where the non-negative integer $\nu$ is the quantum number for the radial excitation and $L_\nu^\lambda$ is an associated Laguerre polynomial.
The eigenenergy is given by
\begin{equation}
  E = 1 + 2\nu + \lambda. \label{eq:eigenenergy}
\end{equation}
Notice that $\lambda$ is not necessarily integral unless $\toyg=0$ or $\toyg\to\infty$.

\subsection{The angular equation}

Integrating Eq.~\eqref{eq:angular-eigenvalue-problem} in an infinitesimal angle around one of the lines $\phi_n$, we see that the derivative of $\Phi(\phi)$ must have a discontinuity of
\begin{equation}
 \lim_{\epsilon \to 0_+} \left( \left.\od{\Phi}{\phi}\right|_{\phi_n+\epsilon} - \left.\od{\Phi}{\phi}\right|_{\phi_n-\epsilon} \right) = \toyg \Phi(\phi_n). \label{eq:kink-condition}
\end{equation}

\begin{figure}
  \includegraphics{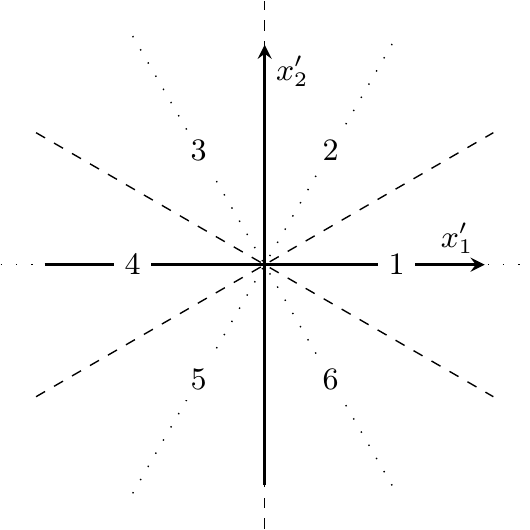}
  \caption{Configuration space in Jacobi coordinates. In hyperspherical coordinates, $\phi$ is the angle to the horizontal axis and $\rho$ is the distance to the origin. Each of the six regions is labeled by its number $n$ and separated from the others by the $\phi_n$ interaction lines (dashed). The dotted lines bisect each region and indicate further reflection symmetries due to relative parity.}
  \label{fig:configuration-space}
\end{figure}

In a region between any two of the $\phi_n$ lines, the problem is that of a free particle on a ring. Enumerate the regions as follows: Region 1 is for $-\frac\pi6<\phi<\frac\pi6$, region 2 for $\frac\pi6<\phi<\frac{3\pi}6$ etc.; see Fig.~\ref{fig:configuration-space}.
In the $n$-th region, we write the wave function as
\begin{equation}
  \Phi(\phi) = a_n e^{i\lambda\phi} + b_n e^{-i\lambda\phi} \label{eq:angular-wave-function}
\end{equation}
for constant coefficients $a_n$ and $b_n$.

The operator $\Lambda_\toyg$ has a discrete rotational symmetry under the transformation $\phi\mapsto\phi+\pi/3$.
It is also symmetric under reflections in the interaction lines (i.e., $\phi=\phi_n$) as well as in the lines halfway between two interaction lines.
The full symmetry group of $\Lambda_\toyg$ is that of a regular hexagon, the dihedral group $\mathrm{D}_6$~\cite{harshman2012}. It is isomorphic to the three-dimensional crystal point group denoted $\mathrm{C}_{6v}$ in Schoenflies notation~\cite{hamermesh}.

We look for solutions of Eq.~\eqref{eq:angular-eigenvalue-problem} among the eigenstates of the operator $C_6$ corresponding to rotations by $\pi/3$; $C_6 \Phi = e^{im\frac{\pi}{3}} \Phi$ for integer $m$.
From this we see that
\begin{equation}
  a_{n+1} = e^{i(m-\lambda)\frac{\pi}{3}} a_n, \quad b_{n+1} = e^{i(m+\lambda)\frac{\pi}{3}} b_n. \label{eq:symmetry-condition}
\end{equation}

Wave function continuity combined with Eqs.~\eqref{eq:kink-condition} and \eqref{eq:symmetry-condition} gives us the relation
\begin{equation}
  \f b_1/{a_1} = \f{2\lambda}/{\toyg} \left( \sin\!\big(\lambda\f\pi/3\big) - \sin\!\big(m\f\pi/3\big) \right) - \cos\!\big(\lambda\f\pi/3\big), \label{eq:B_n-A_n-ratio}
\end{equation}
together with the quantization condition for $\lambda$:
\begin{equation}
  \cos\!\big(\lambda\frac\pi3\big) + \frac{\toyg}{2\lambda} \sin\!\big(\lambda\frac\pi3\big) = \cos\!\big(m\frac\pi3\big). \label{eq:spectrum}
\end{equation}
We notice that the present problem is very similar to that of a one-dimensional particle in a Dirac-comb potential.
Indeed, the dispersion relation of the latter is identical to Eq.~\eqref{eq:spectrum}.
The Dirac-comb problem is, however, often analyzed in the context of solid state physics, e.g.\ Ref.~\onlinecite{griffiths}, where it gives rise to band structure, but we emphasize that this is not the case for our model since here the `lattice' consists of only 6 sites, which is not nearly enough to make a quasi-continuum.

\begin{figure}
  \includegraphics{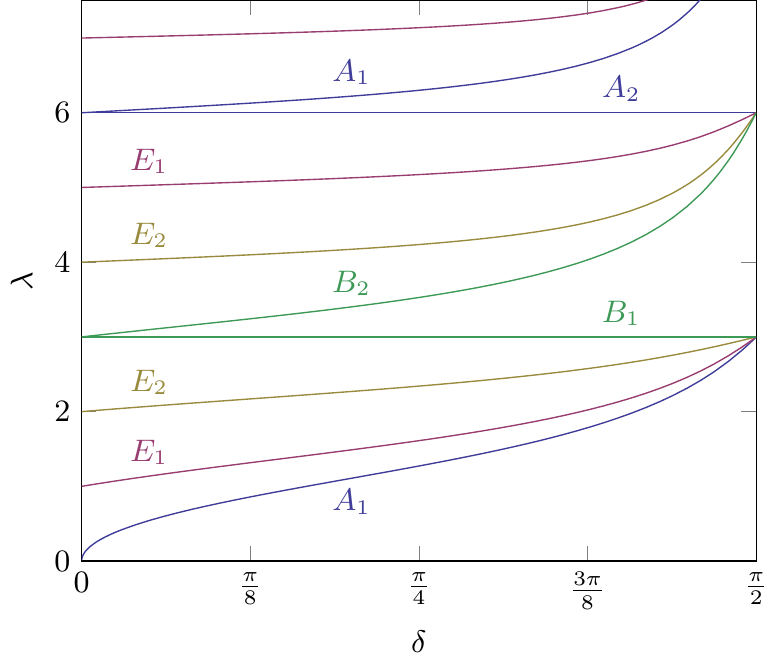}
  \caption{Lower part of the spectrum due to angular excitations. Recall the relation between the energy $E$ and the angular quantum number $\lambda$, cf.~Eq.~\eqref{eq:eigenenergy}. The colors indicate the $m$ quantum number: $|m|=0,6,12,\dotsc$ (blue), $|m|=1,5,7,\dotsc$ (magenta), $|m|=2,4,8,\dotsc$ (gold), $|m|=3,9,15,\dotsc$ (green). Each curve is labeled by the Mulliken symbol of the $\mathrm{D}_{6}$ representation to which it belongs.}
  \label{fig:spectrum}
\end{figure}

\begin{figure*}
  \includegraphics{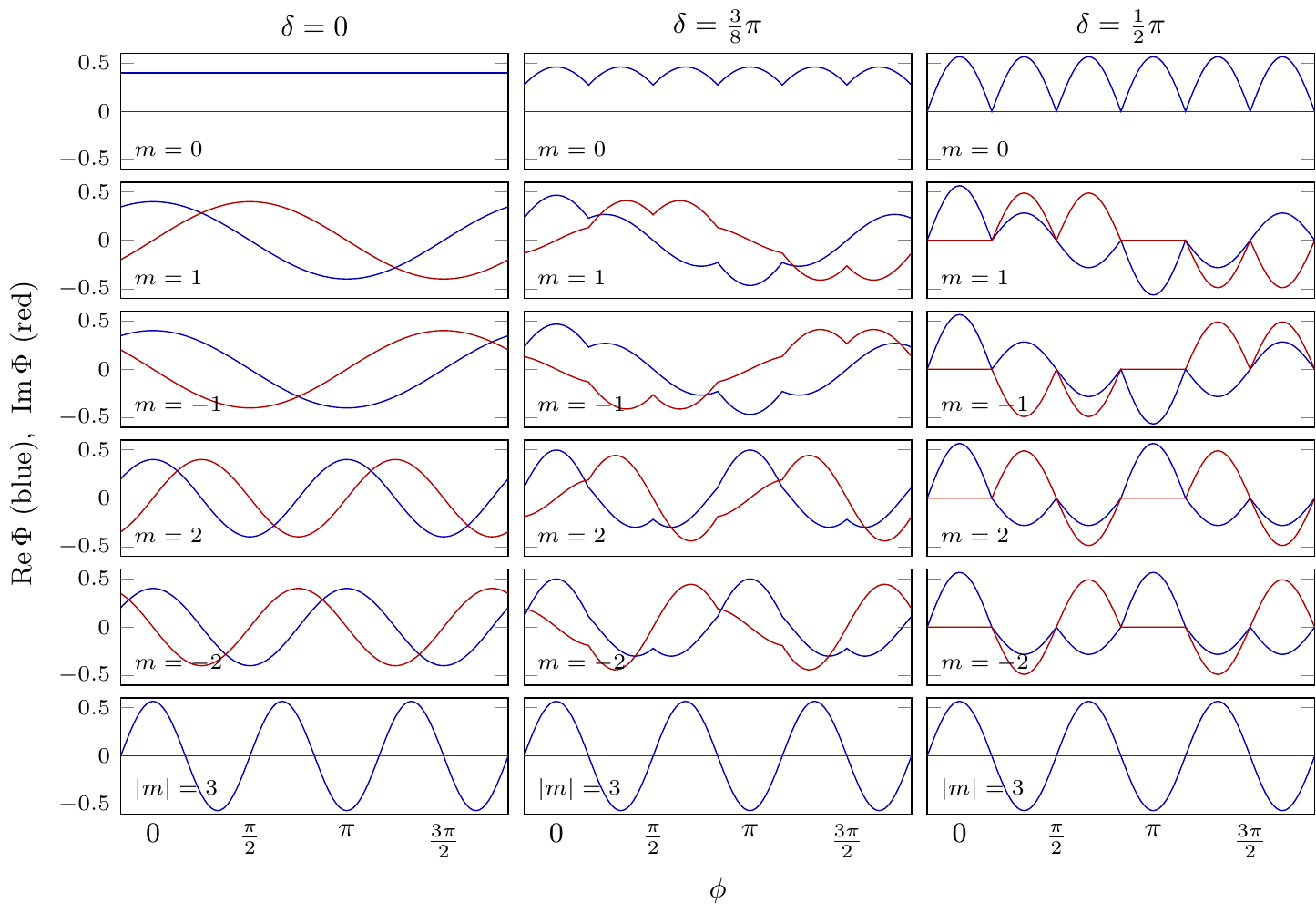}
  \caption{
    The normalized angular wave function $\Phi(\phi)$ plotted against $\phi$ from $-\frac\pi6$ to $\frac{11\pi}{6}$ for varying $\delta$ and $m$.
    The plotted states belong to the lowest strong-interaction multiplet, i.e., they all have $\lambda\to3$ as $\toyg\to\infty$.
    The constant $a_1$ in Eq.~\eqref{eq:angular-wave-function} is taken to be real and positive.
    The left column is the non-interacting limit, the right column is the strongly interacting limit, and the middle column is an example with intermediate interaction strength.
 }
  \label{fig:wave-function}
\end{figure*}

Define the phase $\delta\in[0,\pi/2]$ such that $\tan(\delta)=\toyg/2$. 
In terms of $\delta$, Eq.~\eqref{eq:spectrum} is
\begin{equation}
  \tan(\delta) \sin\!\big(\lambda \frac\pi3\big) = \lambda \left( \cos\!\big(m\frac\pi3\big) - \cos\!\big(\lambda\frac\pi3\big) \right). \label{eq:spectrum-delta}
\end{equation}
The lower part of the spectrum is plotted in Fig.~\ref{fig:spectrum}.
We confirm that for no interactions, the spectrum reduces to that of the free harmonic oscillator, while for $\delta=\pi/2$, it is equivalent to a unitary gas.
Examples of wave functions for $\delta=0$, $\delta=3 \pi/8$, and $\delta=\pi/2$ are given in Fig.~\ref{fig:wave-function}.

\subsection{Symmetries and identical particles}

Amongst the elements of the symmetry group $\mathrm{D}_{6}$ is the operator $\sigma_d$ whose action is to perform a reflection in the $x_1'$-axis, that is, the transformation $\phi \mapsto -\phi$. If $C_6 \Phi = e^{im\frac\pi3} \Phi$, then
\begin{equation}
  C_6 (\sigma_d \Phi) = e^{-im\frac\pi3} (\sigma_d \Phi). \label{eq:m-plus-minus}
\end{equation}

The irreducible representations of the symmetry group $\mathrm{D}_{6}$ are one- or two-dimensional.
Members of the one-dimensional irreducible representations are eigenstates of all the elements of the group, including $\sigma_d$.
So for these states, $|m|$ must be divisible by $3$, such that $e^{im\pi}=e^{-im\pi}$.
Members of the two-dimensional irreducible representations, on the other hand, cannot be eigenstates of both $C_6$ and $\sigma_{d}$.

According to Eq.~\eqref{eq:m-plus-minus}, $\sigma_{d}$ connects states transforming as $e^{im\frac\pi3}$ and $e^{-im\frac\pi3}$ under $C_6$.
So two states with $m=\pm|m|$ (where $m$ is not divisible by $3$) belong to one of the two-dimensional irreducible representations of $\mathrm{D}_{6}$ and are degenerate.

At the limit of no interaction ($\toyg=0$), the angular quantum number is $\lambda=|m|$.
Only the ground state is non-degenerate while every excited state is two-fold degenerate.
Upon introducing a non-zero interaction, $\toyg>0$, the $\mathrm{U}(2)$ symmetry of the free harmonic oscillator in relative configuration space is broken, and the degeneracy of two states characterized by the same rotation number $|m|=3(j+1)$, but belonging to different one-dimensional irreducible representations, is lifted.

Choosing $a_1$ (and thereby also $b_1$) to be real, we have,
\begin{equation}
  \sigma_d (\real \Phi) = \real\Phi, \quad
  \sigma_d (\imag \Phi) = -\imag\Phi,
\end{equation}
so the real and imaginary parts of $\Phi$ are eigenstates of $\sigma_d$.
(When $|m|$ is divisible by 3, either the real part or the imaginary part is identically zero.)
This is confirmed by inspection of Fig.~\ref{fig:wave-function}.

Interchange of two particles corresponds to a reflection in one of the three $\phi_n$ lines.
The $\mathrm{D}_{6}$ group elements responsible for these operations are given by
\begin{equation}
  \sigma_v = \sigma_d C_6^3, \quad
  \sigma_{v'} = \sigma_d C_6, \quad\text{and}\quad
  \sigma_{v''} = C_6 \sigma_d,
\end{equation}
for the particle permutations $(12)$, $(23)$ and $(31)$, respectively \cite{harshman2012}.

We see that $|m|=6,12,\dotsc$ ($|m|=3,9,15,\dotsc$) states with $\sigma_d=-1$ ($+1$) are antisymmetric with respect to interchange of any two particles and, hence, must vanish at every point of interaction; $\Phi(\phi_n)=0$.
They belong to the antisymmetric, irreducible representation of $\mathrm{D}_{6}$ denoted $A_2$ ($B_1$) in Mulliken symbols.
The $A_2$ and $B_1$ states are entirely unaffected by the interaction.
This is exemplified by the wave functions plotted in the bottom row of Fig.~\ref{fig:wave-function}.
The antisymmetric states are also easily identifiable as flat lines in the spectrum Fig.~\ref{fig:spectrum}.

The states that are symmetric with respect to particle exchange belong to $A_1$ for $|m|=0,6,12,\dotsc$ and $B_2$ for $|m|=3,9,15,\dotsc$.
The two-fold degenerate states having $m=0\pm1,6\pm1,12\pm1,\dotsc$ ($m=3\pm1,9\pm1,15\pm1,\dotsc$) belong to the two-dimensional representation $E_1$ ($E_2$) \footnote{We follow the numbering of the irreducible representations used in Ref.~\onlinecite{harshman2012}}.

In the above, we have assumed that the three particles are different.
If on the other hand, some particles are identical, the number of allowed states is reduced.
For example, if two particles are identical while the third is different---a so-called $2+1$ system---the number of states is halved such that the spectrum is degenerate only at $\toyg\to\infty$, where each multiplet is three-fold degenerate.
For $2+1$ fermions, the $A_1$ and $B_2$ states are no longer allowed while $E_1$ and $E_2$ reduce to one-dimensional representations.
For $2+1$ bosons, it is the $A_2$ and $B_1$ states that are forbidden.

If all three particles are identical bosons, only the symmetric representations $A_1$ and $B_2$ are allowed, while if they are identical fermions, only the antisymmetric  $A_2$ and $B_1$ are allowed.

\section{Comparison to three-body contact-interaction model and Calogero model}

In this section we compare the Goldilocks model and the contact-interaction model for three particles. For the contact-interaction model, the same Jacobi angular coordinates can be used and the interaction takes the form
\begin{equation}
V_g = \frac{g}{\sqrt{2}\rho} \sum_{n=1}^6 \delta(\phi-\phi_n).
\end{equation}
As noted before, this potential is no longer separable in Jacobi angular coordinates and there do not exist exact solutions. The spectrum for arbitrary $g$ can be approximated by a variety of schemes, including correlated gaussians~\cite{PhysRevLett.111.045302}, exact diagonalization~\cite{harshman2012,PhysRevLett.111.045302}, analytic approximations based on the exact two-body solutions~\cite{PhysRevLett.108.045301, barfknecht_correlation_2016}, and variational methods~\cite{andersen_interpolatory_2016}. We treat the weak and strong interaction limits using perturbation theory from exact solutions in the next two subsections.

\subsection{Weak interactions}

First we consider the weak interaction limit of the Goldilocks model. To probe the spectrum at weak but non-zero interactions, we differentiate \eqref{eq:spectrum} with respect to $\delta$ and isolate the derivative:
\begin{equation}
  \od{\lambda}{\delta} = \frac{ \sec^2(\delta) \sin(\lambda\frac\pi3) }{ \cos(m\frac\pi3) + \lambda \frac\pi3 \sin(\lambda\frac\pi3) - (1 + \frac\pi3\tan(\delta)) \cos(\lambda\frac\pi3) }. \label{eq:lambda-deriv}
\end{equation}
We see from \eqref{eq:lambda-deriv} and \eqref{eq:eigenenergy} that for small $\delta$ the shift in energy is
\begin{equation}
  \od{E}{\delta}\approx \frac{3}{\pi}\frac{ \sin(\lambda\frac\pi3) }{ \lambda \sin(\lambda\frac\pi3) - \delta\cos(m\frac\pi3) }, \label{eq:lambda-deriv-weak}
\end{equation}
having used that $\cos(\lambda\frac\pi3) \to \cos(m\frac\pi3)$.
An exception is when $\sin(\lambda\frac\pi3)\to0$ as $\delta\to0$, and then we conclude that
\begin{equation}
  \od{\lambda}{\delta}\bigg|_{\toyg=0}
  = \frac{3}{|m|\pi}. \label{eq:toy-model:b-slope-weak-interactions}
\end{equation}
This shows that the non-interacting ground state of the Goldilocks model is extremely sensitive to the interaction, and this result is confirmed by inspection of Fig.~\ref{fig:spectrum}.
Otherwise if $m=3,6,9,\dotsc$, we insert the Taylor expansion $\lambda\approx|m|+\od{\lambda}{\delta}\delta$ into \eqref{eq:lambda-deriv-weak} to obtain
\begin{equation}
  \left( |m| \frac\pi3 \od{\lambda}{\delta} - 2 \right) \od{\lambda}{\delta} = 0,
\end{equation}
having solutions $\dif E/\!\dif\delta=0$ and $\dif E/\!\dif\delta=6/(\pi|m|)$.
From our previous considerations, we know that the former solution holds for the antisymmetric representations $A_2$ and $B_1$.
The latter solution must apply to the $A_1$ and $B_2$ representations.

For comparison, the weak interaction limit of the contact interaction can be calculated using first order perturbation theory. Using the methods of Ref.~\onlinecite{harshman2012}, one calculates
\begin{eqnarray}
\left.\od{E}{g}\right|_{g=0} \!\!\!\!\!\! &=& \frac{A(|m|)}{\sqrt{2}}\int_0^\infty d\rho R^{|m|}_\nu(\rho)^2 \\
&=& A(|m|) \frac{\Gamma(\nu + 1/2)\Gamma(|m| + 1/2)}{\nu!(|m|)!\sqrt{2\pi}} \nonumber\\
&& {} \times {}_3F_2(-\nu,|m| +1/2, 1/2;-\nu+1/2,|m| + 1; 1),\nonumber
\end{eqnarray}
The factor $A(|m|)$ comes from the angular integral and $A(|m|)= 3/\pi$ for $m=0$ and $|m|>0$ except for multiples of three $|m|=3 j $. For $\lambda = 3j>0$ it is either $6/\pi$ (bosonic) or $0$ (fermionic). Note that for $\nu=0$, the hypergeometric function takes the value ${}_3F_2(0,|m| +1/2, 1/2;1/2,|m| + 1; 1) = 1$.

Two key differences between the Golidlocks model and the contact-interaction model are that for the contact-interaction model (1) the slope depends on the radial quantum number $\nu$ and (2) the ground state is no longer as sensitive to the perturbation. Instead of a divergence, it has a finite slope $\dif{E}/\!\dif{g}=2/\sqrt{2\pi}$. 

\subsection{Unitary limit and near unitary limit}

The unitary limit of the contact-interaction model ($1/g = 0$) and the Goldilocks model ($1/\toyg=0$) are equivalent. Further, they coincide with the Calogero model when $\gamma = 0$. As discussed in the next section, this case is maximally superintegrable with five independent integrals of motion with three in involution. The particles are impenetrable, the configuration space becomes disconnected into six ordering sectors, and for distinguishable particles their order becomes a dynamic invariant.

The unitary limit is exactly (algebraically) solvable using the Bose-Fermi mapping~\cite{girardeau_relationship_1960} and its generalization to particles with spin~\cite{PhysRevLett.99.230402, deuretzbacher_exact_2008, guan_exact_2009,fang_exact_2011}. There is a six-fold degenerate level at the unitary limit for every totally antisymmetric solution of the non-interacting problem (assuming distinguishable particles). These six levels can be reduced into one totally symmetric state, two two-dimensional eigenspaces of mixed symmetry, and one totally antisymmetric state that is the same as the free fermionic state. Each of these six-fold levels can be associated with three quantum numbers: center-of-mass excitation $n$, relative radial excitation $\nu$, and relative angular momentum $\lambda = 3(j+1)$, where $n$, $\nu$ and $j$ are non-negative integers. The energy is $(n+ 2\nu + \lambda + 3/2)$ and the degeneracy is six times the number of ways $n$, $\nu$ and $j$ can be chosen to add up to the same energy.

For the Goldilocks model in the near unitary limit of strong but finite interactions (i.e., $\delta$ close to $\pi/2$), the particles are no longer impenetrable. To calculate the first-order energy shift, we can differentiate the energy quantization condition
\begin{equation}
  \od{\lambda}{\delta} \sim -\frac3\pi\sec(\delta)\tan\!\big(\lambda\frac\pi3\big),
\end{equation}
meaning that
\begin{align}
  \frac\pi3 \od{E}{\delta}\bigg|_{\frac1\toyg=0} 
  &= \lambda \left( (-1)^{\lambda+1} \cos\!\big(m\frac\pi3\big) + 1 \right) \\
  &\in \{ 2\lambda, \frac32\lambda, \frac12\lambda, 0 \}. \nonumber
\end{align}
Interestingly, first order perturbation theory for the unitary limit of the contact-interaction gives the same results~\cite{volosniev_strongly_2014, PhysRevA.90.013611, levinsen_strong-coupling_2015}. Evidently the spectrum of the Goldilocks model and the Calogero model are indistinguishable at first order perturbation from the unitary limit for three particles.

Although the contact interaction does not require renormalization for arbitrary interaction strengths in one dimension, starting from the unitary limit and calculating the second-order perturbation of the energy (or first order perturbation of the wave function)  does require renormalization~\cite{sen_perturbation_1999, gharashi_one-dimensional_2015}. One way to understand this is that the wave functions of the energy eigenstates of the contact-interaction model have nodes on the coincidence angles $\phi_n$. Therefore, a naive attempt to construct first-order perturbative wave functions (which do take on non-zero values on the coincidence angles) and from them to calculate the second-order energy shift is bound to fail. However for the Goldilocks model, second-order and higher-order energy shifts can be calculated without renormalization through a Taylor expansion of Eq.~\eqref{eq:spectrum-delta}, demonstrating its possible usefulness in this regime.

As further evidence for this, note that for the Goldilocks model, a first-order perturbation in energy from $E_{\infty}$ (at $\toyg\to\infty$) to $E$ changes the wave function coefficients as
\begin{equation}
  \frac{b_1}{a_1} \simeq (-1)^{\lambda+1} \left( 1 + \frac\pi3 \frac{\sin\big(m\f\pi/3\big)}{(-1)^{\lambda} - \cos\big(m\f\pi/3\big)} (E_{\infty} - E) \right), \label{eq:perturbed-wave-function}
\end{equation}
where $\lambda$ is angular quantum number for the zeroth-order solution.
We notice that the interaction strength $\toyg$ does not explicitly appear in the above equation.
If one inserts the first-order energy shift for the contact-interactions model as $E_{\infty} - E$ in Eq.~\eqref{eq:perturbed-wave-function}, a basis is obtained that may be used to perform a diagonalization of the contact-interactions Hamiltonian in.
The equivalence between the Goldilocks model and the contact-interactions model in energy near the unitary limit suggests that the obtained basis might be a good basis for analytical approximation schemes.

\section{Symmetry, Separability and Integrability for $N>3$}

We have shown that the Goldilocks model for three particles is solvable by separation of variables. For general $0<\toyg<\infty$, solutions are \emph{analytic} but they are not \emph{exact}, where exact solvability means algebraic expressions for the energy and wave functions expressed as polynomials times the ground state~\cite{tempesta_exact_2001, post_families_2012}. Two state labels $n$ and $\nu$ are non-negative integers, and the other state label $\lambda$ is found by solving a transcendental equation. All states are either non-degenerate or two-fold degenerate, assuming distinguishable particles.

However, when $\toyg=0$ or $\infty$ there are exact solutions. All three state labels can be arranged as non-negative integers and the energy spectrum has different degeneracy patterns. In fact, these two limiting cases are \emph{maximally superintegrable}, having five algebraically independent integrals of the motion, or in this context operators defined as continuous transformations on phase space that commute with the Hamiltonian. Understanding these exact solutions, and how these results extend for $N>3$, requires going beyond separability. The next few sections look at the integrability and symmetry of the Goldilocks model.

\subsection{Integrals of motion}

Three integrals of motion for the three-body system are the total Hamiltonian $H_\toyg$, the relative Hamiltonian $H_\text{rel}$ and the angular operator $\Lambda_\toyg$. These integrals are realized by operators that are algebraically independent operators in pair-wise involution, and these operators generate continuous transformations of phase space. This establishes (Liouvillian) integrability for $N=3$. Additionally, there is another operator that commutes with the total Hamiltonian: the total angular operator $\Lambda_\text{tot}$, defined as
\begin{equation}\label{eq:Ltot}
\Lambda_\text{tot} = L_\text{tot}^2 + \sqrt{2}\toyg \rho \sum_{\langle i,j \rangle} \delta(x_i - x_j),
\end{equation}
where $L_\text{tot}^2$ is the normal three-dimensional angular momentum squared operator in configuration space. Similarly, the operator $\Lambda_\toyg$ can be expressed
\begin{equation}\label{eq:Lrel}
\Lambda_\toyg = L_\text{rel}^2 + \sqrt{2}\toyg \rho \sum_{\langle i,j \rangle} \delta(x_i - x_j),
\end{equation}
where $L_\text{rel}^2 = -\partial^2/\partial{\phi}^2$. Note that the operator $\Lambda_\text{tot}$ does not commute with the other two integrals of motion $H_\text{rel}$ and $\Lambda_\toyg$. The extra integral of motion can be associated with the separability of the Hamiltonian in spherical coordinates as well as cylindrical coordinates~\footnote{The classical Goldilocks model is additionally separable in oblate and prolate spherical coordinate, based on the results of Evans~\cite{PhysRevA.41.5666}.}. Therefore, the three-body Goldilocks model is \emph{minimally superintegrable} in the terminology of Evans~\cite{PhysRevA.41.5666}.

These same four operators $H_\toyg$, $H_\text{rel}$, $\Lambda_\toyg$, and $\Lambda_\text{tot}$ can be generalized to any $N$ using the well-known higher-dimensional generalization of angular momentum. However, there are still only three integrals in involution and four total, therefore not enough to integrate the $N=4$ case or higher.

The expressions (\ref{eq:Ltot}) and (\ref{eq:Lrel}) for the total and relative angular operators make it clear than in the limit $\toyg =0$ these operators are the standard angular momentum squared operators in total and relative configuration space. This case corresponds to the $N$-dimensional isotropic harmonic oscillator, which is maximally superintegrable and massively multi-separable~\cite{doi:10.1063/1.1704724}. There are multiple ways to choose the additional  algebraically independent quadratic operators that realize the missing $2N-5$ integrals of motion~\cite{doi:10.1063/1.1704724, PhysRevA.41.5666, moshinsky_harmonic_1996}.

In the limit $\toyg \to \infty$, the $N$-body Goldilocks model also corresponds to the $\gamma =0$ limit of the $N$-body Calogero-Moser model. This limiting model is also maximally superintegrable but is provably not separable for $N \geq 4$~\cite{doi:10.2991/jnmp.2005.12.s1.43}. Note that in this limit the four integrals of motion are no longer bounded on the Hilbert space of Lebesgue square-integrable functions on configuration space. However, one can restrict the domain to only those functions that have zero support on the manifold defined by the $N(N-1)/2$ coincidence planes. The other integrals of motion required for maximal superintegrability are not quadratic~\cite{gonera_calogero_199}. In future work, we plan to explicitly construct these integrals and use them to determine which integrals are preserved in the near-unitary limit. The model in the near unitary limit can be mapped onto an integrable spin chain~\cite{volosniev_strongly_2014, PhysRevA.90.013611}. Perhaps this method will also explain why the ansatz for the spin chain coupling coefficients presented in Ref.~\onlinecite{levinsen_strong-coupling_2015} is  surprisingly effective.

\subsection{Symmetry and degeneracy}

In principle, every energy level of a Hamiltonian should correspond to an irreducible unitary representation of the kinematic symmetry group of the Hamiltonian. By kinematic symmetry group, we mean a group that acts on configuration space or phase space and is represented by unitary operators that commute with the Hamiltonian~\footnote{Note that some systems have degeneracies which cannot be formulated under the framework of kinematic symmetries, like the Pythagorean degeneracies of $N$ free particles in an infinite square well. See Ref.~\onlinecite{PhysRevA.95.053616} for a recent discussion that points out some of the limits of this definition.}.

For the $N$-body Goldilocks model with general $0<\toyg<\infty$, the  kinematic symmetries of the model contain the following subgroup~\cite{Harshman2016}:
\begin{equation}\label{eq:ksub}
\mathrm{U}(1) \times \mathrm{O}(1) \times S_N  
\end{equation}
The first factor describes rotations that mix the separable center-of-mass position and momentum coordinates in phase space, or equivalently the symmetry of re-phasing the center-of-mass creation and annihilation operators. The second factor is inversion by relative parity; for three particles this is the rotation $\phi \mapsto \phi + \pi$. The third factor comes from the exchange symmetry of identical particles. The first two factors are both Abelian groups with one dimensional irreducible representations (irreps), so any degeneracies must correspond to the dimensions of the irreps of $S_N$ (see Ref.~\onlinecite{hamermesh} for a detailed description of $S_N$ irreps). For three particles, this agrees with previous results: for each relative parity there are one dimensional irreps for states symmetric or antisymmetric under exchange and two dimensional irreps for mixed symmetry states. This should extend to all $N$ without any change, leading, for example, to one-, two-, and three-fold degeneracies for four distinguishable particles and to one-, four-, five-, and six-fold degeneracies for $N=5$.

Additionally, the kinematic group must also contain the three independent, one-parameter groups generated by $H_\toyg$, $H_\text{rel}$, and $\Lambda_\toyg$. These groups also have one-dimensional irreps and therefore do not change the degeneracy.

Note that the contact interaction shares the same kinematic symmetry subgroup (\ref{eq:ksub}). The loss of separability removes $\Lambda_\toyg$ and its one-parameter subgroup from the kinematic symmetry, but they have the same structure of degeneracies. For a discussion of the limiting cases of $\toyg =0$ and $\toyg \to \infty$ which coincide with the same  limits of the contact-interaction model, see Ref.~\onlinecite{Harshman2016}.

\subsection{Dynamical $\SO(2,1)$ symmetry}

Finally, we want to comment on the dynamical (or hidden) $\SO(2,1)$ symmetry~\cite{werner-castin-pra} (or equivalently  $\mathrm{SL}(2, \mathbb{R})$ symmetry~\cite{gonera_calogero_199}). Define the operators
\begin{equation}
  W_{\pm} = \f1/2 \left( H_\toyg - \rho^2 \pm \left(\frac{N-1}{2} + \rho \pd{}{\rho}\right) \right)
\end{equation}
as in Ref.~\onlinecite{werner-castin-pra}.
These operators satisfy the commutation relations
\begin{equation}
  [H_\toyg, W_\pm] = \pm2 W_\pm \quad\text{and}\quad
  [W_-,W_+] = H_\toyg,
\end{equation}
and generate a hidden $\SO(2,1)$ symmetry in the system. In other words, they do not commute with the total Hamiltonian, but they do map energy eigenstates into other energy eigenstates, like ladder operators. The action of the operators is to increase or decrease the radial quantum number $\nu$ by one unit:
\begin{equation}
  \begin{aligned}
    W_+ R_\nu^\lambda(\rho) &= \sqrt{(\nu+1)(\nu+\lambda+1)} R_{\nu+1}^\lambda(\rho), \\
    W_- R_\nu^\lambda(\rho) &= \sqrt{\nu(\nu+\lambda)} R_{\nu-1}^\lambda(\rho).
  \end{aligned}
\end{equation}
In the $N$-particle case, the solution to the relative radial equation is
\begin{equation}
  R_\nu^\lambda(\rho) = \sqrt{\frac{2\nu!}{\Gamma(\nu+\lambda+\frac{N-1}{2})}} \rho^{\lambda} e^{-\rho^2/2} L_\nu^{\lambda+{(N-3)}/{2}}(\rho^2),
\end{equation}
if the angular equation is taken to be
\begin{equation}
  \Lambda_\toyg \Phi = \lambda(\lambda + N - 3) \Phi.
\end{equation}
The corresponding Casimir operator is
\begin{equation}
  C = H_\toyg^2 - 2 (W_+ W_- + W_- W_+) = \Lambda_\toyg - 1.
\end{equation}

Notice that the above considerations regarding a hidden $\SO(2,1)$ symmetry are not specific to the Goldilocks model. A similar analysis applies to all systems---having any number of particles or dimensions---as long as the Hamiltonian is separable in relative hyperspherical coordinates. This separation is possible for any quadratic external field, but only for interaction with the correct scaling.

\subsection{Comparison with no harmonic trap}

For the sake of completeness, we briefly consider the case of no external trapping field. 
Informally, this can be thought of as the zero-frequency limit of the trapped model. 
However, we cannot take the zero-frequency limit of the Goldilocks Hamiltonian (\ref{ham:toy}) (or the contact-interaction  or Calogero Hamiltonians (\ref{ham:con}) and (\ref{ham:cal})), because the trap frequency has been absorbed into natural units length scale. When there is no trap, the physically relevant length scale for all three models is set by the interaction strength parameter.

The separable center-of-mass motion is now unbounded. Mathematically, instead of the harmonic symmetry $\mathrm{U}(1)$ in (\ref{eq:ksub}), the kinematic symmetry is the Euclidean group of one-dimensional translations and reflections $\mathrm{E}_1$ (not to be confused with $E_1$, the notation for the two-dimensional irrep of $\mathrm{D}_6$). The center-of-mass momentum is an integral of motion, but we note that the corresponding generator does not have proper eigenvectors in the Hilbert space, only generalized (Dirac) eigenkets. Except for the zero-energy state, the irreps of $\mathrm{E}_1$ are two-dimensional. In other words, there are two states with the same energy that are mixed by reflections. Further, the untrapped model has the dynamical symmetry of Galilean transformations in one-dimension $\mathrm{G}_1$. This group contains the spatial symmetry $\mathrm{E}_1$ but is extended by one-dimensional boosts.

In the relative motion, the absence of the trapping potential changes the nature of the hyperradial solutions, but preserves the hyperradial separability. This separability means the three-body Goldilocks model is still minimally superintegrable and exactly solvable for $N=3$, but neither for $N>3$. The Calogero model retains its maximal superintegrability for all $N$. Most interestingly, for the contact-interaction model, the lack of the trapping potential makes the scattering interaction diffractionless. The Hamiltonian is integrable and solvable via the Bethe ansatz~\cite{sutherland_beautiful_2004}.

\section{Conclusion}

The argument of this article went from solving a three-body model, comparing it to known results for a related model, and then analyzing the separability, symmetry and integrability. Of course, the idea behind and motivation for the article was reversed: use symmetry to identify solvable models, see how they compare to physical models of known interest, and then solve them to gain insight. The goal of this avenue of research is to have a toolbox for identifying when a model can be solved and, when it cannot be solved, to have a method for finding nearby solvable models. Then these nearby models used to interrogate the few-body physics through direct application if relevant and possible, or through analytic and numerical extensions like variational methods,  perturbation methods, and analytical approximation techniques. In light of this motivation, the Goldilocks model should be a useful tool in the one-dimensional few-body toolbox.

An extension of this work is explicitly constructing the integrals of motion at the limiting case and seeing how perturbations from the limits break some of these integrals and preserve others. Because one-dimensional atomic gases have been considered as possible working material for quantum sensors, quantum simulators, and quantum information processing devices, quantifying the robustness of integrability under perturbations could have direct application.

Finally, a natural question is to ask what happens in higher dimensions. The contact-interaction models in higher dimensions require regularization or renormalization for rigorous treatment~\cite{busch_two_1998, 1367-2630-14-5-053037, 1751-8121-45-20-205302}. The modified two-dimensional contact interaction is already relative hyperradial-hyperangular separable without any further changes, leading to the special nature of the two-dimensional solutions~\cite{busch_two_1998, 1751-8121-45-20-205302}.  The three-dimensional (modified) contact interaction requires multiplication (not division) by $\rho$ in order to become separable. In this case, each pairwise interaction in the $N$-body interaction potential is reduced or screened by the close presence of other particles, i.e.\ the opposite of what takes place in one dimension. There may be interesting applications of this potential, but the one additional integral of motion that comes from separability is not going to make as much of a difference to solvability or integrability in the $6N$ phase space.

\section*{Acknowledgements}

This research was supported by the Danish Council for Independent Research and the DFF Sapere Aude program, and by the Aarhus University Research Foundation. Part of this work is contained in the Master's thesis of M.E.S.A.  Thanks to our colleagues A.S. Jensen and D.~Blume for conversations about the (then-unnamed) Goldilocks model.

\end{document}